\documentclass[conference]{IEEEtran}
\usepackage{amsmath,amssymb,amsfonts}
\usepackage[ruled,vlined]{algorithm2e}
\usepackage{graphicx}
\usepackage{biblatex}
\def\BibTeX{{\rm B\kern-.05em{\sc i\kern-.025em b}\kern-.08em
    T\kern-.1667em\lower.7ex\hbox{E}\kern-.125emX}}
\usepackage{dblfloatfix} 

\ifCLASSOPTIONcompsoc
    \usepackage[caption=false, font=normalsize, labelfont=sf, textfont=sf]{subfig}
\else
\usepackage[caption=false, font=footnotesize]{subfig}
\fi

\ifCLASSINFOpdf
\else
\fi
\begin{document}
%
% paper title
% Titles are generally capitalized except for words such as a, an, and, as,
% at, but, by, for, in, nor, of, on, or, the, to and up, which are usually
% not capitalized unless they are the first or last word of the title.
% Linebreaks \\ can be used within to get better formatting as desired.
% Do not put math or special symbols in the title.

\title{Megha: Decentralized Global Fair Scheduling for Federated Clusters}

% Things to do:

% 2. Check consistency in vocabulary
% 3. Refine related works
% 4. Describe results better
% 5. Add summary of results to introduction
% 6. Run it through grammarly
% 7. Resize and position images 
% 8. Check references and labelling
% 9. Submit!!!!!

% author names and affiliations
% use a multiple column layout for up to three different
% affiliations

\author{\IEEEauthorblockN{Meghana Thiyyakat}
\IEEEauthorblockA{\textit{PES University}\\
Bangalore, India \\
meghanathiyyakat@pesu.pes.edu}
\and
\IEEEauthorblockN{Subramaniam Kalambur }
\IEEEauthorblockA{\textit{PES University}\\
Bangalore, India \\
subramaniamkv@pes.edu}
\and
% \linebreakand
\IEEEauthorblockN{ Dinkar Sitaram}
\IEEEauthorblockA{\textit{Cloud Computing Innovation Council of India}\\
dinkar@ccici.in}
}

% conference papers do not typically use \thanks and this command
% is locked out in conference mode. If really needed, such as for
% the acknowledgment of grants, issue a \IEEEoverridecommandlockouts
% after \documentclass

% for over three affiliations, or if they all won't fit within the width
% of the page (and note that there is less available width in this regard for
% compsoc conferences compared to traditional conferences), use this
% alternative format:
% 
%\author{\IEEEauthorblockN{Michael Shell\IEEEauthorrefmark{1},
%Homer Simpson\IEEEauthorrefmark{2},
%James Kirk\IEEEauthorrefmark{3}, 
%Montgomery Scott\IEEEauthorrefmark{3} and
%Eldon Tyrell\IEEEauthorrefmark{4}}
%\IEEEauthorblockA{\IEEEauthorrefmark{1}School of Electrical and Computer Engineering\\
%Georgia Institute of Technology,
%Atlanta, Georgia 30332--0250\\ Email: see http://www.michaelshell.org/contact.html}
%\IEEEauthorblockA{\IEEEauthorrefmark{2}Twentieth Century Fox, Springfield, USA\\
%Email: homer@thesimpsons.com}
%\IEEEauthorblockA{\IEEEauthorrefmark{3}Starfleet Academy, San Francisco, California 96678-2391\\
%Telephone: (800) 555--1212, Fax: (888) 555--1212}
%\IEEEauthorblockA{\IEEEauthorrefmark{4}Tyrell Inc., 123 Replicant Street, Los Angeles, California 90210--4321}}

% use for special paper notices
%\IEEEspecialpapernotice{(Invited Paper)}

% make the title area
\maketitle

% As a general rule, do not put math, special symbols or citations
% in the abstract

\begin{abstract}
%With production workloads becoming more complex and diverse, data center schedulers today need to meet the increasing scheduling demands of these workloads while guaranteeing fairness to users and high utilization of the data center's resources. 
Increasing scale and heterogeneity in data centers have led to the development of federated clusters such as KubeFed, Hydra, and Pigeon, that federate individual data center clusters.
In our work, we introduce Megha, a novel decentralized resource management framework for such federated clusters. Megha employs flexible logical partitioning of clusters to distribute its scheduling load, ensuring that the requirements of the workload are satisfied with very low scheduling overheads. It uses a distributed global scheduler that does not rely on a centralized data store but, instead, works with eventual consistency, unlike other schedulers that use a tiered architecture or rely on centralized databases.
 Our experiments with Megha show that it can schedule tasks taking into account fairness and placement constraints with low resource allocation times - in the order of tens of milliseconds.
 %which is 6x-14x lower than the current state-of-the-art data center schedulers.
\end{abstract}
% no keywords

% For peer review papers, you can put extra information on the cover
% page as needed:
% \ifCLASSOPTIONpeerreview
% \begin{center} \bfseries EDICS Category: 3-BBND \end{center}
% \fi
%
% For peerreview papers, this IEEEtran command inserts a page break and
% creates the second title. It will be ignored for other modes.
\IEEEpeerreviewmaketitle

\section{Introduction}
% no \IEEEPARstart
%   \subsubsection{Need for a single shareable pool of resources.\footnote{Hydra}} 
   To meet the growing demands of scale, organizations are investing in data centers with hundreds of thousands of machines. These data centers are further divided into clusters based on application or utility. High utilization of this infrastructure evaluates to higher returns on investment. To achieve high utilization, the resources in a data center must be presented as a shareable unified pool to its users, and fine-grained allocation from this pool must be supported. 
  
  Due to the heterogeneity and complexity prevalent in recent workloads, scheduling and resource allocation are non-trivial problems. To manage a large number of worker nodes efficiently, scheduling frameworks are moving towards federated architectures wherein a top-level scheduler makes high-level scheduling decisions which are enforced by independent cluster schedulers. Hydra \cite{curino2019hydra} and Pigeon\cite{wang2019pigeon} are two such schedulers that have embraced the federated architecture.

    In our paper, we introduce Megha, a decentralized global scheduler that uses eventual consistency and flexible partitioning to overcome the limitations of existing schedulers. Megha has a federated architecture and divides the data center into smaller clusters, each of which is controlled by a Local Master (LM). It achieves high scheduling throughput by distributing the load to multiple top-level Global Masters (GMs). Each GM uses an eventually-consistent global view of the system's resources to make scheduling decisions. The LMs validate the decisions of the GM and deploy the tasks on the worker nodes of the clusters handled by them. In our work, we also demonstrate how global fairness policies can be enforced with eventual consistency.
    
    Our experiments were conducted with real-world production traces to evaluate Megha's task allocation speed and compare it with centralized and distributed scheduling approaches. The results show that Megha can achieve a median allocation time comparable to that of Sparrow \cite{ousterhout2013sparrow}, a random-sampling based distributed scheduler. Megha also reports two orders of magnitude better 99th percentile allocation times than Sparrow. The experiments also demonstrate how task placement constraints can be satisfied with eventually consistent global state while reporting lower allocation times than the centralized approach.
    
     \textbf{Our Contributions}
     Our main contributions are listed below. We:
     \begin{itemize}
         \item define allocation time and break it down into its multiple components. We also analyze the contribution of each component towards allocation time in different scheduling approaches.
         \item introduce the design of a novel decentralized framework that uses eventual consistency and dynamic partitioning to overcome the limitations of existing approaches.
         \item explain how fairness can be implemented in a decentralized scheduler.
         \item demonstrate the improvement in allocation time of tasks scheduled using our approach when compared to a centralized scheduler, and a distributed scheduler(Sparrow).
         \item explore the behavior of the framework under different parameter settings.
     \end{itemize}
     
     The rest of the paper is organized as follows. Section~\ref{sec:relwork} discusses related work. Section ~\ref{sec:alloc_time} defines allocation time and discusses its different components. Section~\ref{sec:megha} gives an overview of the architecture of the framework and the fairness algorithm used. Section~\ref{sec:eval} describes the evaluation methodology. Section~\ref{sec:results} presents the results of the experiments.  Section~\ref{sec:conclusion} concludes the findings of the paper and Section~\ref{sec:future} briefly describes future enhancements to our work.  

\section{Related Work}\label{sec:relwork}

\subsection{Heterogeneity in workloads}
To achieve better utilization of data center resources, jobs from multiple applications are co-located on the same data center. Recent studies \cite{cheng2018characterizing,jassas2018failure,reiss2012heterogeneity} have analysed production workload traces published by companies such as Google \cite{wilkes2019google} and Alibaba\cite{alibaba18}. An analysis\cite{lu2017imbalance} of the Alibaba cluster trace published in 2017 exposes the imbalance in the CPU and memory consumed by the tasks in the trace. This variation in the proportions in which the resource types are consumed advocates the need for fine-grained, independent resource allocation of each resource type. According to a study \cite{guo2019limits} of the cluster trace published in 2018, Alibaba's workload consists of jobs from over 9K online services. Each online service has a different SLO in terms of tail latency. The study also discusses the variation in the characteristics of batch jobs and service jobs in terms of resources consumed and the runtime duration. Apart from the heterogeneity present within a workload, workloads also vary across datacenters. Due to this variation, it is recommended \cite{amvrosiadis2018diversity} that multiple cluster traces be used while evaluating scheduling solutions. In our work, we use three production traces to evaluate Megha.

\subsection{Placement Constraints}
From routine maintenance and upgrades in the data center, applications tend to run on heterogeneous hardware spanning multiple generations, running different software versions. Due to the diversity in the infrastructure, tasks have placement constraints that limit the machines on which the tasks can be run. Resource constraints dictate task placement based on resource availability, whereas placement constraints are based on the software and hardware configurations of machines, such as the OS kernel version and the clock speed of the CPU. In their work, Sharma et al. \cite{sharma2011modeling}, discuss the impact of placement constraints on scheduling. According to the authors, taking into account placement constraints increases the delay in scheduling tasks by a factor of 2 to 6. However, these constraints cannot be ignored as they play an important role in ensuring optimal job performance. Nearly 50\% of all tasks in Google's workloads have simple non-combinatorial constraints. 
The results of our experiments with Megha are consistent with their findings that the scheduling overhead is very small when the load on the system is low. However, when the load increases, the unavailability of the resources that satisfy the tasks' constraints translates to longer task wait times, and hence, higher allocation times.

\subsection{Scheduling Approaches}

Schedulers can be broadly classified into centralized, distributed, hybrid, and hierarchical schedulers. Centralized schedulers, such as YARN \cite{vavilapalli2013apache}, can make optimal decisions because they have a global view of the system. However, as previous studies have shown \cite{curino2019hydra} as the size of the data center increases, the processing and collection of large amounts of state lead to the formation of a bottleneck due to which these schedulers are unable to achieve the required scheduling throughput.

Distributed architectures such as Sparrow \cite{ousterhout2013sparrow} rely on probability to find machines with available resources. For each task, Sparrow performs random sampling and inserts probes into the worker queues to pick the machine with the lowest queuing time. The scheduling overhead in this approach is insignificant. But, when the data center is highly loaded, the probability of picking an available machine is very low. Therefore, distributed schedulers that rely on random sampling perform poorly under high utilization conditions.

The architecture that most closely resembles Megha's is that of Pigeon \cite{wang2019pigeon}. Pigeon divides the data center into smaller groups, each of which is managed by a master. Distributed schedulers assign tasks in a job evenly across masters. In Megha, the GM is analogous to the distributed scheduler and the LM is analogous to the master. However, there are three main differences between the two architectures. Firstly, Pigeon does not perform fine-grained resource allocation. It divides its workers into fixed resource encapsulations called slots. Such coarse-grained allocation will lead to fragmentation and wastage of resources. Secondly, Pigeon uses weighted fair queuing and reservation to avoid long job starvation and to ensure low short job latency, respectively. Megha does not differentiate between job types. However, due to its decentralized nature, Megha achieves low allocation times and ensures against head-of-line blocking and starvation of any class of jobs. Thirdly, the groups in Pigeon once formed are fixed, that is, there is no scope for flexibility in the partitioning scheme. Megha allows tasks to be scheduled anywhere in the data center by using flexible partitioning. This allows better utilization of the data center's resources, especially in the presence of placement constraints which restricts the number of eligible machines on which a task can run.

Hydra \cite{curino2019hydra} is a federated resource management framework, created by Microsoft as Apollo's \cite{boutin2014apollo} successor, to support the high scheduling decisions per second required by the workloads deployed on their data centers. Hydra divides its large cluster into smaller YARN sub-clusters. Each sub-cluster, therefore, has a Resource Manager (RM), that decides the task placement and performs the allocation of resources. Hydra introduces a component, called the AM-RM Proxy that allows tasks to span multiple sub-clusters. In Megha, the Global Master is analogous to the AM-RM Proxy while the Local Masters plays the role of the RMs. However, because each Megha GM maintains a (possibly stale) local copy of the global view of the entire system, it doesn't need to contact the LMs every time a task request arrives. Hence, Megha can achieve lower median task allocation time (in the order of tens of milliseconds), than those reported by Hydra (2-3 seconds).

Hybrid schedulers such as Mercury \cite{karanasos2015mercury}, Hawk \cite{delgado2015hawk}, and Eagle \cite{delgado2016job} use two sets of schedulers - a centralized scheduler for tasks that need resource guarantees and distributed schedulers for latency-sensitive tasks. These schedulers, however, suffer from the same problems as the distributed schedulers mentioned earlier. Moreover, due to the lack of coordination between the two sets of schedulers, global fairness policies cannot be implemented.

PCSsampler \cite{hao2017pcssampler} extends the random sampling-based approach by caching state information received in response to probes. The usefulness of the cached state is highly dependent on the workers probed. In large clusters under high loads, finding a worker with available slots may require multiple rounds of probing leading to high scheduling overheads. PCSsampler also does not take into account fairness. Megha's GMs, on the other hand, collect partial state information about an entire LM cluster after every launch request. The GMs also update their stored global state using periodic heartbeats from LMs. Megha ensures all its decisions are optimal by having the LMs validate them.

Other works \cite{delimitrou2013paragon,delimitrou2014quasar,romero2018mage, yang2013bubble} optimize the performance of tasks by using machine learning to improve the task placement quality taking into account resource and task heterogeneity and the interference posed by co-located tasks.

\section{Allocation Time}\label{sec:alloc_time}

Short jobs from user-facing applications such as web searches and other queries typically run for hundreds of milliseconds. Since the quality of the users' experience depends on the response time of these jobs, they must be scheduled with negligible overheads. Such latency-sensitive tasks cannot tolerate resource allocation times greater than tens of milliseconds \cite{ousterhout2013sparrow}. 

We define resource allocation time as the time it takes for a task to be allocated resources in a worker node and begin its execution. We measure resource allocation time as:
 \begin{equation}
 Allocation\_Time = Task\_Start - Task\_Arrival\_Time    
 \end{equation}
  \(Task\_Start\) is the time at which the task starts executing on a worker node and \(Task\_Arrival\_Time\) is the time at which the task is inserted into the framework's request queue. Resource Allocation Time can be split further into 3 components:
\begin{multline}
Allocation\_Time= Framework\_Queuing\_Delay \\+ Processing\_Delay + Worker\_Queuing\_Delay 
\\+ Communication\_Delay
\end{multline}
\(Framework\_Queuing\_Delay\) is the time the task spends in the framework's request queue waiting for processing. When the scheduling throughput of the framework is low, this delay can contribute to large resource allocation times. \(Processing\_Delay\) is the time taken by the framework to find a worker node that satisfies the requirements of a task. The framework must take into account resource constraints, and additionally,  placement constraints, while finding an eligible worker node. Centralized scheduling frameworks need to process the entire global state of the system to find a worker node for a task. Simultaneously, they also need to ensure their state is kept up-to-date. While the consistent global view ensures that the scheduling decisions made are optimal, the increase in the number of workers leads to an increase in the amount of processing required, and consequently the creation of a bottleneck resulting in higher framework queuing delays. 
\(Worker\_Queuing\_Delay\) is the time spent by the task in the chosen worker node's queue while it waits for resources to be allocated to it. Distributed scheduling frameworks like Sparrow \cite{ousterhout2013sparrow} and the distributed schedulers in hybrid frameworks such as Eagle\cite{delgado2016job}, have an insignificant  \(Framework\_Queuing\_Delay\) component due to the high throughput of the frameworks and small \(Processing\_Delay\) component characteristic of random sampling. However, in these frameworks, the \(Worker\_Queuing\_Delay\) component significantly impacts the  \(Allocation\_Time\). Distributed schedulers do not have access to a global view of the system. Therefore, when the system is under high load, the schedulers are unable to find available worker nodes with certainty. This results in the queuing up of tasks at busy worker nodes even while there are available worker nodes in the system.

\(Communication\_Delay\) is the delay introduced due to communication between multiple components in the framework requiring them to send messages over the internal network. This value is dictated by the network speed of the data center and the number and size of the messages sent.

% Formally capture objectives using mathematical notations
% \begin{itemize}
%     \item Must take into account multiple resources while scheduling \footnote{Yaguchi}.
%     \item Fine-grained resource allocation. 
%     \item Scalability - Microsoft requires clusters with 50k nodes. Fleet consists of multiple such clusters. Advantages of large clusters.
%     \item Low scheduling overhead and high throughput - Hydra paper says 40k decisions per second need to be made, Borg requires 10k.
%     \item Placement Constraints - cite Google paper. Give formal definitions of machine constraints, task constraints.
%     \item Jobs spanning clusters - points from Hydra paper
%     \item System utilization?
    % \item Placement Quality - how is it measured? \footnote{Briefly describe in Introduction (contributions). Describe in detail in Results section. Is this needed if we have fairness?}
    % \begin{itemize}
    % \item ? Gang scheduling
    % \item Fairness
    % \item Compare to Sparrow and Eagle
    
    % \item Response time
    % \item \% Placement constraints
% \end{itemize}

\section{Megha's Architecture}\label{sec:megha}

\subsection{Overview}
\begin{figure}[!b]
    \centering

       \includegraphics[width=\linewidth]{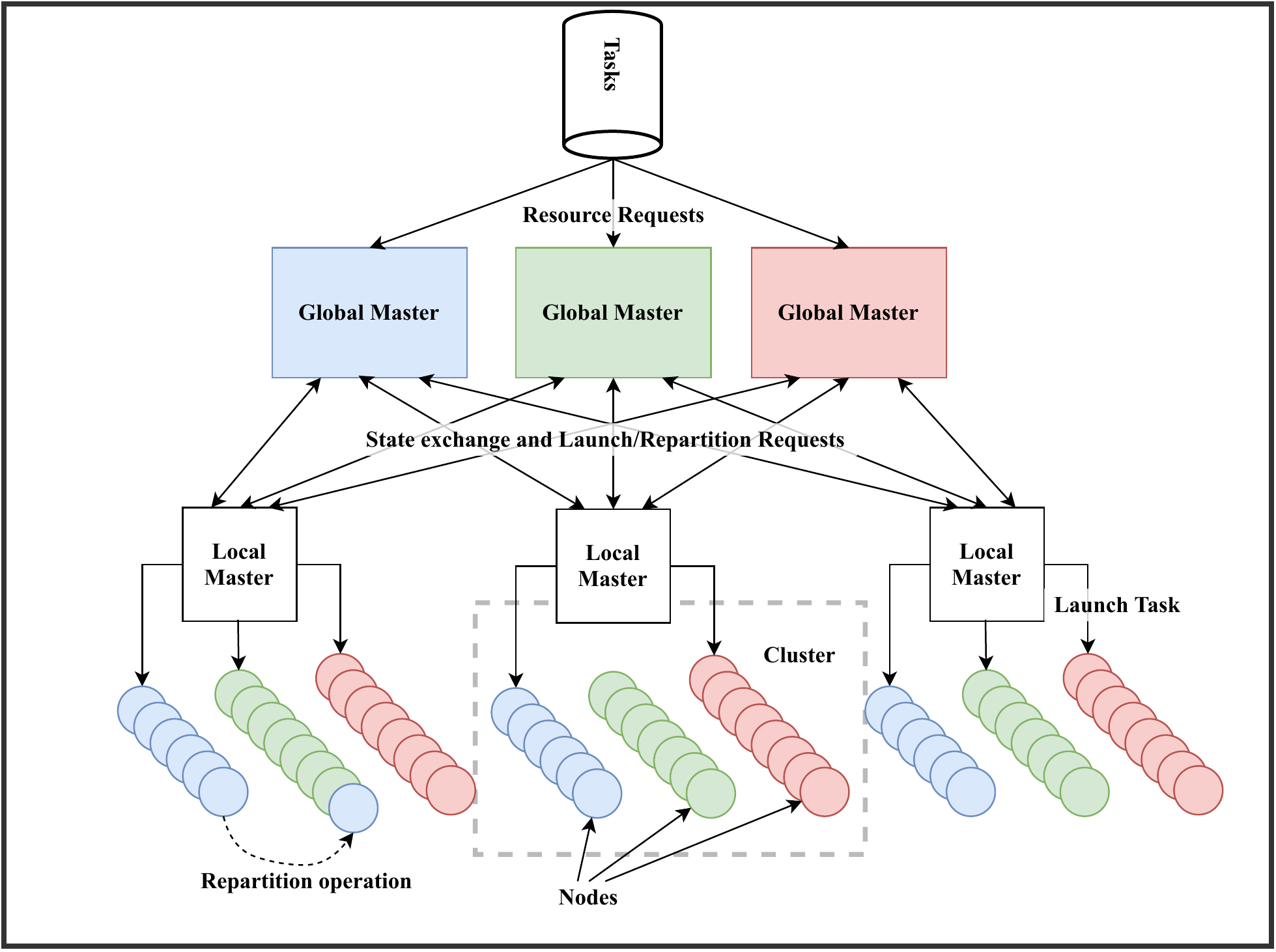}
        
    \caption{Megha's architecture with one user queue}
    \label{fig:arch}
\end{figure}

Megha has a decentralized architecture that combines the superior placement quality of centralized scheduling, with the scalability and low allocation times associated with distributed scheduling. The framework federates multiple smaller clusters in the data center and manages them as a single pool of resources. Each smaller cluster is governed by a Local Master (LM) which is analogous to a centralized resource manager such as YARN. The Local Master performs the typical functions of a resource manager- sends tasks to worker nodes in the cluster, checks for worker heartbeats, collects status information from workers, handles failure recovery of workers, etc. The task placement decision, of which node to launch a task on, are made by Global Masters (GMs) and the task launch requests are sent to the LMs.A depiction of Megha's architecture is shown in Fig. \ref{fig:arch}.

The GMs are independent of each other and do not communicate with one another. The sharing of the resources of the data center amongst them is facilitated by the LMs. Each LM periodically sends the GMs updates on the state of its clusters, enabling the GMs to have a global view of the entire system. However, the global state information maintained by a GM may be stale. The LMs on other hand have a partial view of the system - restricted to their cluster only; but, the state stored by an LM is always consistent and current. Therefore, the LMs are responsible for verifying if the GMs' requests can be satisfied by the true current state of the system. This is important because the resources requested by a GM may have been allocated to another GM. Aside from the periodic updates, the LMs also send updates to the GMs, piggybacking the response to task launch requests made by the GMs.

Each cluster under an LM is divided into logical sub-clusters called \textit{partitions}. The LM assigns each of the GMs one of its partitions. Hence,

\begin{equation}
count(partitions)=count(GMs)
\end{equation}

 In Fig. \ref{fig:arch}, worker nodes in the cluster have been depicted using circles. The worker nodes in a partition have been colored uniformly, and the worker nodes are assigned to the correspondingly colored Global Master. Partitions assigned to a GM are known as its internal partitions. The remaining partitions, which belong to other GMs, are referred to as external partitions. Therefore, each GM has at its disposal the resources of a logical cluster formed by the aggregation of its internal partitions under each of the LMs. That is,

\begin{equation}
GM\_cluster_j=\bigcup_{i=1}^{i=n} P_{ij}  
\end{equation}

where \(n\) is the number of LMs, and \(P_{ij}\) is the partition in \(LM_i\) assigned to the \(GM_j\). 

\subsection{Tasks and Machine Constraints}
 We borrow the concept of placement constraints from previous work \cite{sharma2011modeling}. A machine is said to have a machine constraint if it has the property that satisfies the requirement denoted by the constraint. For example, a machine with a clock speed of 3.1 GHz satisfies the constraint "clock speed greater than 2.5GHz". A task placement constraint, on the other hand, denotes a preference or requirement of the task in terms of the hardware or software configuration of the machine. Therefore, a task, with the task placement constraint mentioned in the example above, can be scheduled on any machine with a clock speed greater than 2.5 GHz. A task can have zero or more placement constraints. A machine that satisfies all the placement constraints of a task is a potential candidate for the task to be scheduled on.
\subsubsection{Representation}Each placement constraint is represented by a unique number in Megha. For every partition, the GM maintains a bit-vector per constraint such that each bit in the vector corresponds to a worker node in the partition. Therefore, the length of the bit-vector is equal to the number of worker nodes in the partition. The value of the bit represents whether the corresponding node satisfies the requirements of the constraint. In our work, we have considered 21 task placement constraints. Thus, a task or machine can have any combination of the 21 constraints. The GM maintains 21 bit-vectors for every partition in the system - both internal and external.

\subsubsection{Processing Task Requests} 
A task request, \(T = (R,PC)\), made to a GM consists of the resource requirements of the task  \(R=(R_1, R_2 ... R_n\)), and a list of task placement constraints  \(PC= (PC_1, PC_2 ... PC_k\)). Here, \(n\) is the number of resource types, such that, \(1  \leq n  \leq r \), where \(r\) is the total number of resource types that Megha is aware of and \(R_i\) represents the number of units of resource \(i\) that the task requires.  \(k\) is the number of task placement constraints, such that, \( 0\leq k \leq m\), where \(m\) is the number of machine constraints available in the system.

A worker node in the system is of the form \(N= (K,M)\), where \(K\) is the representation of the resources available in the worker node and \(M\) is the list of task placement constraints that the worker node satisfies, known as its machine constraints. On receiving a task request, the GM, uses its global state and Megha's matching algorithm, to pick a worker node from one of its internal partitions based on the task's placement constraints. That is, if a task has a list of placement constraints \(PC\), such that \(PC = (PC_1, PC_2 ... PC_k\)), the GM searches for a worker node with machine constraints \(M = (M_1,M_2 ...M_l)\), such that 
\begin{equation}\label{eq:mc}
M \supseteq PC
\end{equation}

where, a machine constraint \(M_i \in M\), satisfies the task's placement constraint \(PC_i \in PC\). The GM then checks its global state to see if the worker node has sufficient resources available. Therefore, if \(K = (K_1,K_2...K_j)\) is the list of resources available in the worker node \(N\), where \(K_i\) denotes the number of units of resource \(i\), and \(R = (R_1,R_2...R_n)\) is the resource requirement of the task \(T\), then \(N\) is considered an eligible match for \(T\) if

\begin{equation}\label{eq:rc}
  \forall i \in n :   K_i \geq R_i
\end{equation}

 \subsubsection {Match operation} \label{sec:match} When a task request arrives at a GM, it chooses a partition from its internal partitions in a round-robin fashion. It then checks which worker nodes in the partition can satisfy the combination of placement constraints that the task requires. It does this by performing a bit-wise AND operation between all the bit-vectors corresponding to the task's constraints. The position of the positive bits in the resulting vector tells the GM which worker nodes in the partition satisfy all the placement constraints of the task. The GM then loops through the resource availability of the matching worker nodes and checks if any of the worker nodes can satisfy the resource requirements of the task. The match operation's algorithm has been shown in Algorithm \ref{alg:matching}. The GM sends the LM a launch request of the form \((N, T)\) when it encounters a worker node with sufficient resources. The LM confirms the availability of the requested resources and launches the task on the chosen worker node. 
 
 Since the match operation uses a bit representation, its time complexity can be represented as $\mathcal{O}( n . partition\_size )$, where \texttt{n} is the number of task placement constraints. However, since the AND operation is performed either at the word or byte level, depending on the operand size in the CPU architecture, $\mathcal{O}( n . log_{size}partition\_size)$,  would be a more accurate representation of the time complexity, where \(size\) denotes the operand size.

\begin{algorithm}[!t]
    \SetAlgoLined
    \KwIn{ task request: $request$\\ task placement constraints: $constraints^T$\\ constraints matrix: $constraints^N[NUM\_CONSTRAINTS]$}
    
    \Begin{
    \SetKwFunction{Fun}{Function match\_task}
    //initialize to constraint vector of 1s \\
    // size NUM\_CONSTRAINTS \\
    $candidates$=[11...1] \\
    \Fun{$request$, $constraints^T$}{\\
    // bitwise AND of constraint vectors \\
         \ForEach{ $constraint$ $\in$ $constraints^T$}{
        $candidates$=\\$candidates$ $AND$ $constraints^N[constraint]$    
        }
        
        \ForEach {$candidate \in candidates$}
        {
            \uIf{ $candidate == 1$ AND $request \leq candidate.resources$ }
            {
            // match found successfully \\     
                return $candidate$
            }
        }
        // no match found\\
        return $-1$
        
    }
    
   }
     \caption{Pseudo-code for the matching algorithm}
     \label{alg:matching}
\end{algorithm}

\subsubsection{Repartition operation}

When a task request cannot be satisfied by the internal partitions of a GM, the GM looks at the resources in the external partitions. If it finds an eligible worker node for the task, that is, a worker node that satisfies (\ref{eq:mc}) and  (\ref{eq:rc}), it requests the LM to perform a repartition operation. In a repartition operation, the LM temporarily adds the required resources from the eligible worker node into the GM's internal partition. It does so by assigning a logical worker node to the GM's internal partition with just the requested resources and deducting these resources from the actual node's pool of available resources. Therefore, we formally define a repartition operation on a node \(N=(K,M)\) for a task \(T=(R,PC)\) as the creation of a logical node \(N'=(K',M)\), where \(K' = R \), and updating \(N\) such that \(N=(K - R, M)\). The logical node \(N'\) is added to the GM's internal partition in the LM.

\begin{figure}[]
    \centering
       \includegraphics[width=0.85\linewidth]{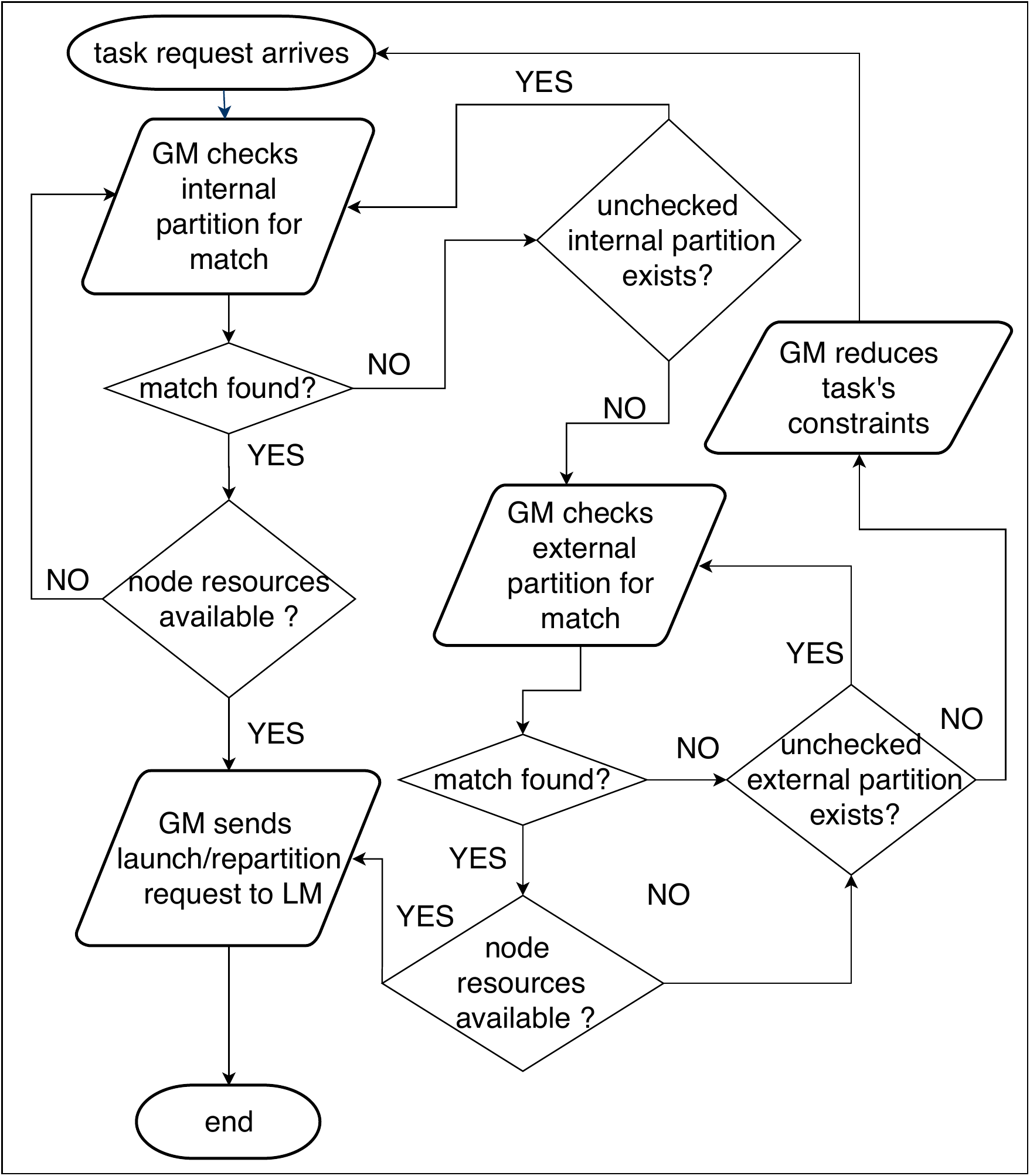}
        
    \caption{Simplified representation of a GM's decision making process}
    \label{fig:flow}
\end{figure}

 This capability of Megha to place tasks anywhere in the data center using the cached global state ensures that all task placement constraints are satisfied with minimal overhead. A simplified representation of the GM's decision-making process is shown in Fig. \ref{fig:flow}. It is important to note that the repartitioning is performed at resource level granularity(number of vCPUs, MB of Memory) and not at the individual worker node level. This increases the probability of the framework finding a match for a task while also promoting the sharing of resources, thereby improving the utilization of the data center. To distribute the load evenly across all LMs, the GM uses round-robin while picking external partitions.
 
While the figure shows the movement of a worker node from one partition to another, it must be noted that this movement is implemented with minimal overhead by changing only the logical mapping of the worker node to a partition. The figure also simplifies the depiction of the repartition operation by showing it at the granularity of an entire worker node. However, Megha is capable of more fine-grained repartitioning, as mentioned earlier. 
 
 If none of the worker nodes in the partition can satisfy both the placement constraints and the resource requirements of the task, the GM moves on to the next partition and repeats the process. This continues until the GM finds an eligible worker node or all the internal partitions have been searched without finding a match. When no match is found, the GM looks at the external partitions. If an eligible worker node is found in one of the external partitions, the GM requests the LM to perform a repartition operation. 

% \subsubsection{Constraint Simplification operation} When a launch request (or repartition request) to an LM cannot be satisfied, the GM continues to search in the remaining partitions. However, when no worker node in the system can satisfy all the placement constraints of the task, the GM performs a Constraint Simplification operation. When a GM is forced to perform a Constraint Simplification operation, it removes the lowest priority constraint from the task's list of constraints and reschedules the task. Megha assumes that each constraint's priority is reflected by the number used to represent the constraint. Therefore, the higher the constraint number, the lower its priority. The Constraint Simplification operation is performed until a worker node satisfying the updated placement constraints of the task can be found in the system. 

\subsubsection{Rescheduling operation} \label{sec:reschedule} When a task's requirements cannot be satisfied by the internal partitions of a GM or by repartitioning, the GM inserts the task request at the end of its task request queue. When the request reaches the front of the queue, the GM treats it like a new request and follows the regular decision-making process as depicted in Fig. \ref{fig:flow}.

\subsubsection{State Inconsistencies} \label{sec:inconsistentcy}As mentioned earlier, the GM’s view of the state of the system, while global, maybe stale or inconsistent. This may cause the GM to make invalid launch requests to the LM requesting resources that are no longer available. There are two scenarios where the GM can make incorrect launch requests. The first is where the requested resources from the GM’s internal partitions have been assigned to another GM as a result of a repartition operation. The second scenario is where the GM makes a repartition request but the requested resources are no longer available in the external partition - either due to the resources having been consumed by the GM in charge or due to a repartition commissioned by a third GM. Both scenarios are a result of an inconsistent or stale global state in the GM. When such a request is made to the LM, the LM responds with a failure message along with the current state of all its partitions. The overhead of state inconsistencies is equal to the sum of the time taken to make a launch request to the LM, and the amount of time the LM takes to check its cluster state and respond.

\subsection{Fairness in Megha}
Megha uses queues to enforce fair sharing. Each user is assigned a queue configured with the user's share of the data center's resources. All tasks from the user's jobs are submitted to this queue. Each queue is mapped to one GM only. However, a single GM can service requests from multiple queues belonging to different users. The processes the task requests from the queues in using round-robin.
To ensure optimal utilization of data center resources, Megha enforces fairness only when there is resource contention. Therefore, the GM does not perform checks when a task request (with resource and placement constraints) can be satisfied by its internal partitions or by a repartition operation. Only when the GM is unable to find an eligible worker node for a task does the fairness algorithm come into play. When a match cannot be found, the GM first checks if the requesting user has already consumed her share of the resources. If not, the GM finds the user with the largest share violation and preempts the corresponding user's tasks such that it yields the required resources. However, if the requesting user has already consumed her share, the task request is re-inserted into the user's queue. The pseudo-code of the fairness algorithm is given in Algorithm \ref{alg:fairness}.

Each GM has consistent state information about its own user queues only and has eventually-consistent (possibly stale) information about other user queues. Since preemption requests are made to preempt tasks belonging to other queues, preemption requests have to be verified by the LM. If the verification fails, the GM finds other eligible tasks for preemption and repeats the process. 
Since the GM has access to the consistent state of its own queues, our approach ensures against any of the queues unfairly consuming excess resources when there is resource contention. When there is resource contention, the GM checks if the user queue has already consumed its share. If it has, the task cannot request for preemption of tasks belonging to other user queues. However, due to the GM's inconsistent state, it may preempt tasks from queues that are no longer in violation of their share. In such a scenario, after the preempted task reaches the front of the task queue again, it is scheduled by the GM as a regular task. If sufficient resources are not available in spite of the queue adhering to its share, the GM can preempt tasks from other queues that are in violation. In addition to the heartbeats it receives, the GM updates its state regarding the resource consumption of the queues each time it interacts with an LM.

\begin{algorithm}[!t]
    \SetAlgoLined
    \KwIn{ task request: $request$}
    
    \Begin{
    \SetKwFunction{Fun}{Function get\_Preempt\_Tasks}
    \Fun{$request$}{
    \\
        \uIf{$share(request.user) < resources\_consumed(user)$}
            {
                return Failure
                % $num\_threads = ceil(row.cpu\_time/row.window) $ 
                % $row.cpu\_time  = row.cpu\_time / num\_threads $
            
                % \For{$i\gets0$ \KwTo $num\_threads$ \KwBy $1$}{
                %   execute  $RUN\_ON\_CPU(row)$ on a new thread
                % }
            
            }
            \Else
            {
                // Check user in decreasing order of violation
                \While{$preempt\_user=get\_next\_user()$}{%
                    $preempt\_tasks=get\_preemption\_tasks(preempt\_user)$
                    \uIf{$preempt\_tasks$ !  empty}{return $preempt\_tasks$
                    }
                        
                }
                return [] // empty list
            } 
    }
    
    \SetKwFunction{args}{Function Main}
    \args{$request$}{
    
    $status=Launch\_Task\_Internally(request)$
    \\
     \uIf{$status != Success $}
            {
                $status=Repartition\_and\_Launch(request)$
                \\
                \uIf{status != Success}
                {
                    $tasks=get\_Preempt\_Tasks(request)$
                    \\
                    \uIf{$tasks$ empty }{
                        Reinsert request in user's task queue
                    }
                    \Else{
                    $Preempt\_and\_Launch(tasks,request)$
                    }
                }
            
            }
       }}
     \caption{Pseudo-code for the fairness algorithm}
     \label{alg:fairness}
\end{algorithm}

\subsection{High Availability}

The Global Masters in Megha are stateless. Each GM's state can be reconstructed on failure from the heartbeats received from the Local Masters. Failure recovery of Local Masters and the worker nodes can be borrowed from traditional approaches. On failure, a worker can be restarted by the LM in charge; the LM must relaunch all tasks previously running on the worker with the help of its active task list. High availability of the LMs can be implemented using an active-standby configuration, such that, both the active and the standby LMs are sent all the updates from the workers and the GMs, but only the active LM enforces the GMs' scheduling decisions.

\section{Evaluation}\label{sec:eval}
To demonstrate that Megha outperforms both distributed and centralized approaches, we compare its decentralized approached with Sparrow and a single-LM-single-GM configuration of Megha which behaves like a centralized scheduler. We study the performance of Megha under different scenarios with different cluster sizes and framework parameters. In all the scenarios, the heartbeat period is set to 10 seconds. 

We have evaluated our framework with workloads derived from production traces. The workloads contain task resource requests, task durations, and task placement constraints. A prototype of the framework was tested on 128 cloud instances and the simulation of the framework was evaluated for data centers of sizes up to 50,000 worker nodes.

An algorithm from previous work \cite{sharma2011modeling} was used to model the heterogeneity of the data center and to assign various machine constraints to the worker nodes in each cluster. Each cluster has been generated with machine constraint probabilities of one of the clusters- A, B, or C, published in the work, chosen at random. Therefore, the clusters under each LM bear different distributions of the machine constraints.

\subsection{Workloads}
Our experiments were conducted using 3 workload samples- \texttt{google\_19}, \texttt{alibaba} and \texttt{cloudera}. They have been derived from the production traces published by Google \cite{wilkes2019google}, Alibaba \cite{alibaba18} and Cloudera\cite{chen2012interactive}. The \texttt{google\_19} workload consists of 4550 tasks, the \texttt{alibaba} workload consists of 8000 tasks and the \texttt{cloudera} workload consists 10173 tasks.

To evaluate the performance of Megha for tasks with placement constraints, we augmented each task in the \texttt{alibaba} and \texttt{google\_19} traces with placement constraints using the methodology, and distributions published in previous work \cite{sharma2011modeling}. We use a sample of the \texttt{cloudera} trace to compare the performance of Sparrow and Megha under different load conditions. We use the \texttt{google\_19} trace sample to demonstrate Megha's global fair scheduling.

The resource requests of the workloads have been scaled down for the experiments with the prototype. The CPU requested by the tasks have been scaled down by a factor of 400, and the memory has been scaled down by a factor of 50. However, the number of tasks, the duration of the tasks, the inter-arrival times, and the task placement constraints remain identical.

\subsection{Simulation}
 For the simulation, we have considered data centers from 1000-50,000 workers. For all data center sizes, the worker nodes were given a configuration of 64 CPU cores and 16GB of RAM. The simulations were run on a 4-core Intel Xeon E5-2683 v4 machine.  We use the publicly available Sparrow simulator in our work. The Sparrow simulator does not consider any overheads other than a constant network delay and queuing delays at the workers. The Megha simulator is a single-node deployment of the prototype. Therefore, unlike the Sparrow simulator, it also takes into account scheduling overheads such as the overheads of message processing and the overheads from the decision-making incurred at each LM, GM, and worker node. The only delay that the simulator does not account for is the delay due to interference from the co-location of tasks. For a fair comparison with Megha, we run the Sparrow simulator with a modification to the network delay parameter to account for message processing overhead. We retain the default value of the network delay (0.5ms) while calculating the overhead for probing. However, we calculate the network overhead of the task information being sent to the worker differently. We replace the default network delay for the operation with the median of the transfer time for task launch messages recorded by Megha since these messages are expected to be the same in both frameworks.
 
\subsection{Prototype}
The prototype was deployed on 128 nodes on the Linode cloud service. Each node had 2 vCPUS and 4GB of RAM. The size of the cluster under each LM was 40 nodes. Each GM and LM ran on a separate node. The Request Generator and RabbitMQ servers were deployed on one node each.

In all scenarios, each worker node in the system has been assigned machine constraints according to the probability distributions and algorithms mentioned earlier.

\subsection{Implementation}
Megha uses HTTP REST APIs for synchronous communication such as launch, repartition, and preemption requests, and RabbitMQ messages for asynchronous updates. Both the simulation and the prototype have been implemented using Python3.8. Gunicorn 20.0.4 has been used as the HTTP server in all the components. RabbitMQ v3.8.5 server was used for communication.

\section{Results}\label{sec:results}

\subsection{Comparison with distributed and Centralized schedulers}

\begin{figure}[!b]
    \centering
      \includegraphics[width=0.8\linewidth]{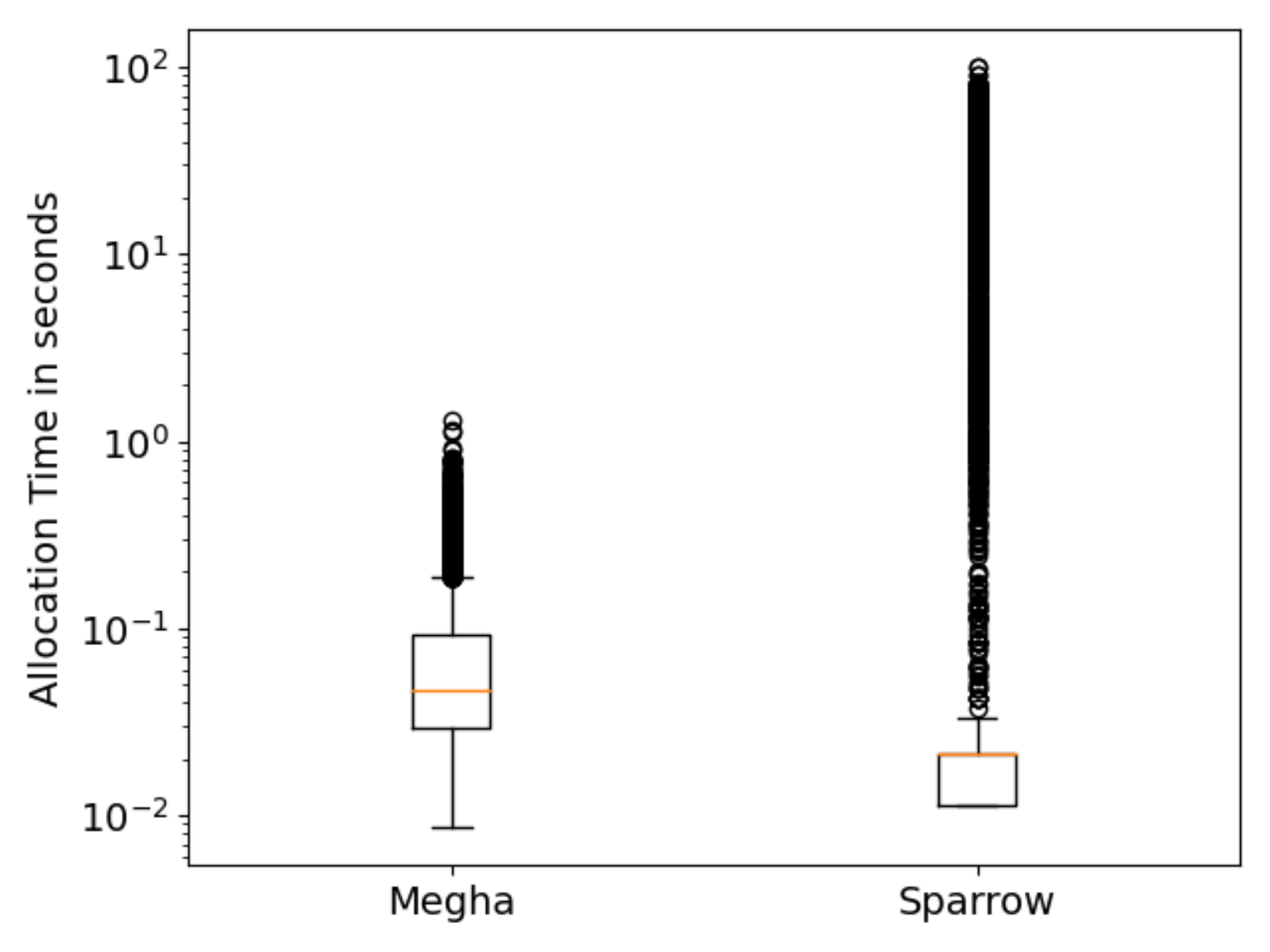}
    \caption{
    Distribution of allocation times reported by Megha and Sparrow}
    \label{fig:dist}
\end{figure}

\begin{figure}[!t]
    \centering
      \includegraphics[width=0.8\linewidth]{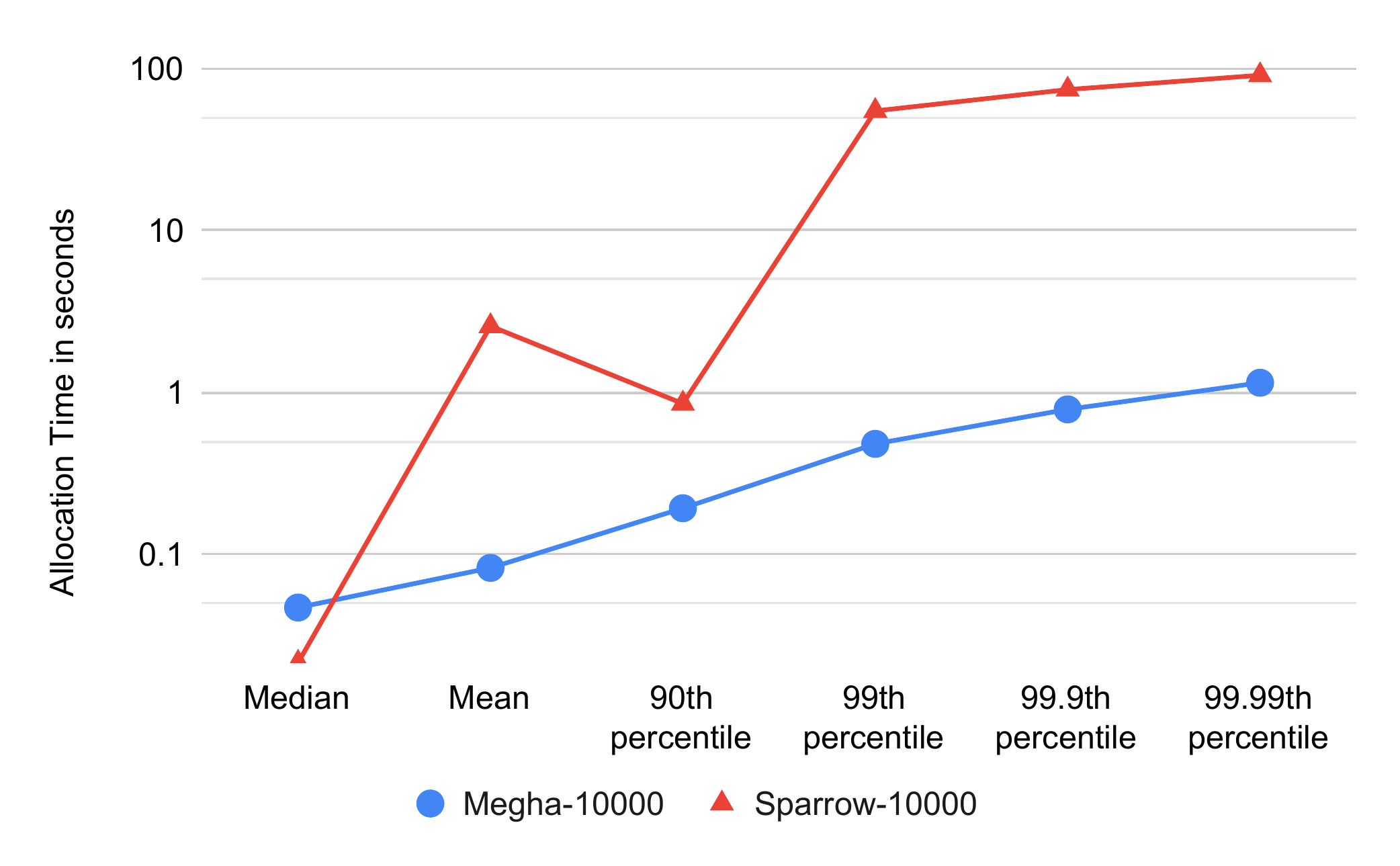}
    \caption{Median, Mean, 90th percentile,99th percentile, 99.9th percentile and 99.99th percentile Allocation time recorded for Megha and Sparrow}\label{fig:sparrow}
\end{figure}

\begin{figure}[!t]
    \centering
       \includegraphics[width=0.8\linewidth]{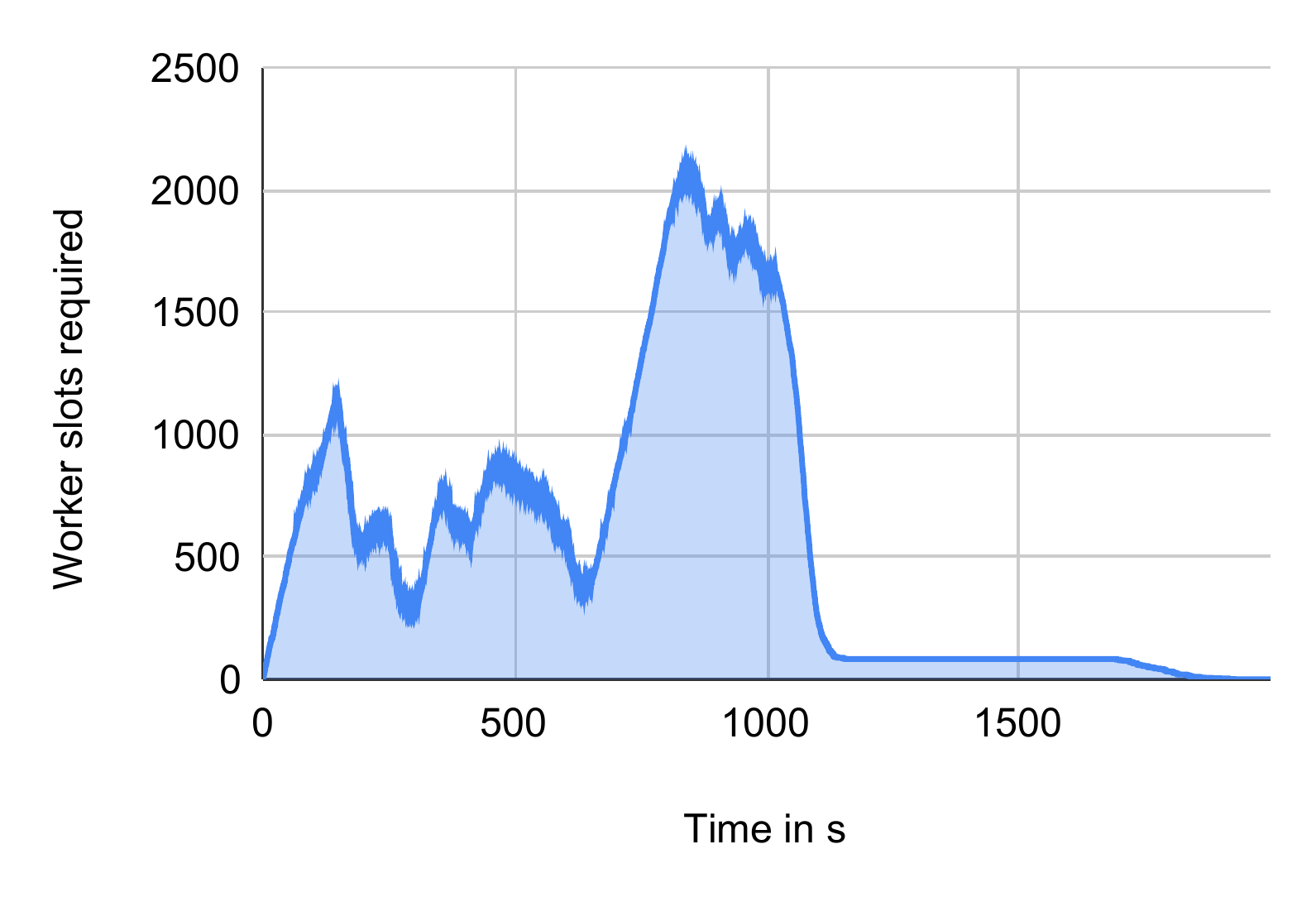}
    \caption{Total number of worker slots required by the workload sample}
    \label{fig:cc}
\end{figure}

We compare the resource allocation times recorded for Sparrow and Megha for 10k workers in Fig. \ref{fig:dist} and \ref{fig:sparrow}. The resource consumption curve of the Cloudera workload is plotted against time in Fig. \ref{fig:cc}. The median allocation time of Sparrow is smaller than Megha's. However, it can be observed in the graphs that Megha's allocation time is consistently less than 1.5s, whereas the allocation times recorded under Sparrow have a large variation and can run into tens of seconds (max:~100s). The long tail in Sparrow's distribution can be attributed to its inability to ascertain which nodes are available using random-sampling.

\begin{figure}[!b]
    \centering
    
       \includegraphics[width=0.9\linewidth]{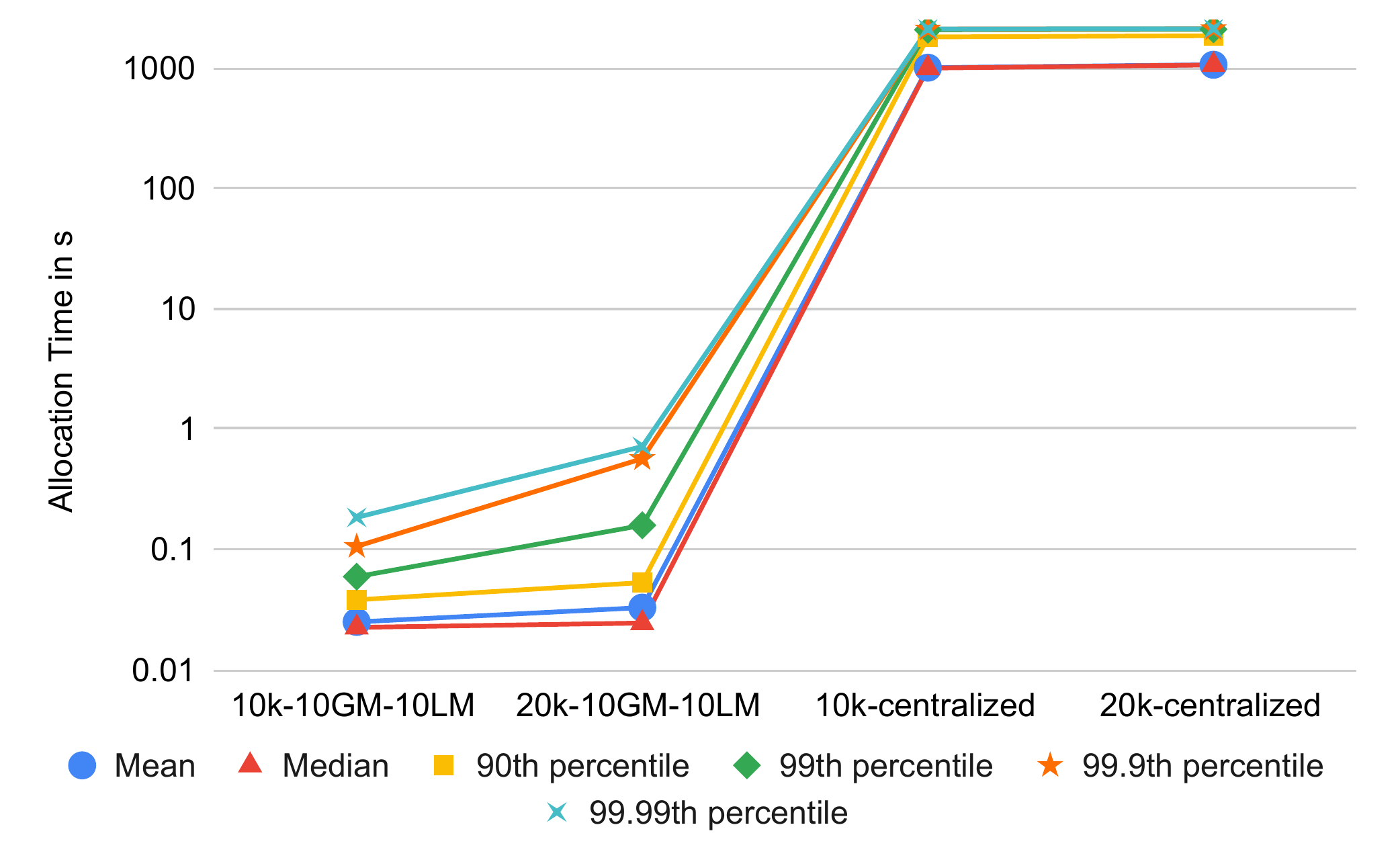}
    \caption{Comparison of allocation time for 10k and 20k worker nodes for centralized and decentralized approaches}
    \label{fig:central}
\end{figure}

We also compare the resource allocation times reported by Megha's decentralized configuration and its centralized configuration (single-LM-single-GM). We use the placement constraints-augmented \texttt{alibaba} trace for this purpose. The results are shown in Fig. \ref{fig:central}. Due to the decrease in parallelism in the GMs and LMs, the centralized case reports a very high allocation time when compared to Megha.

\subsection{Component Delays}

\begin{figure}[!t]
    \centering
    
       \includegraphics[width=0.9\linewidth]{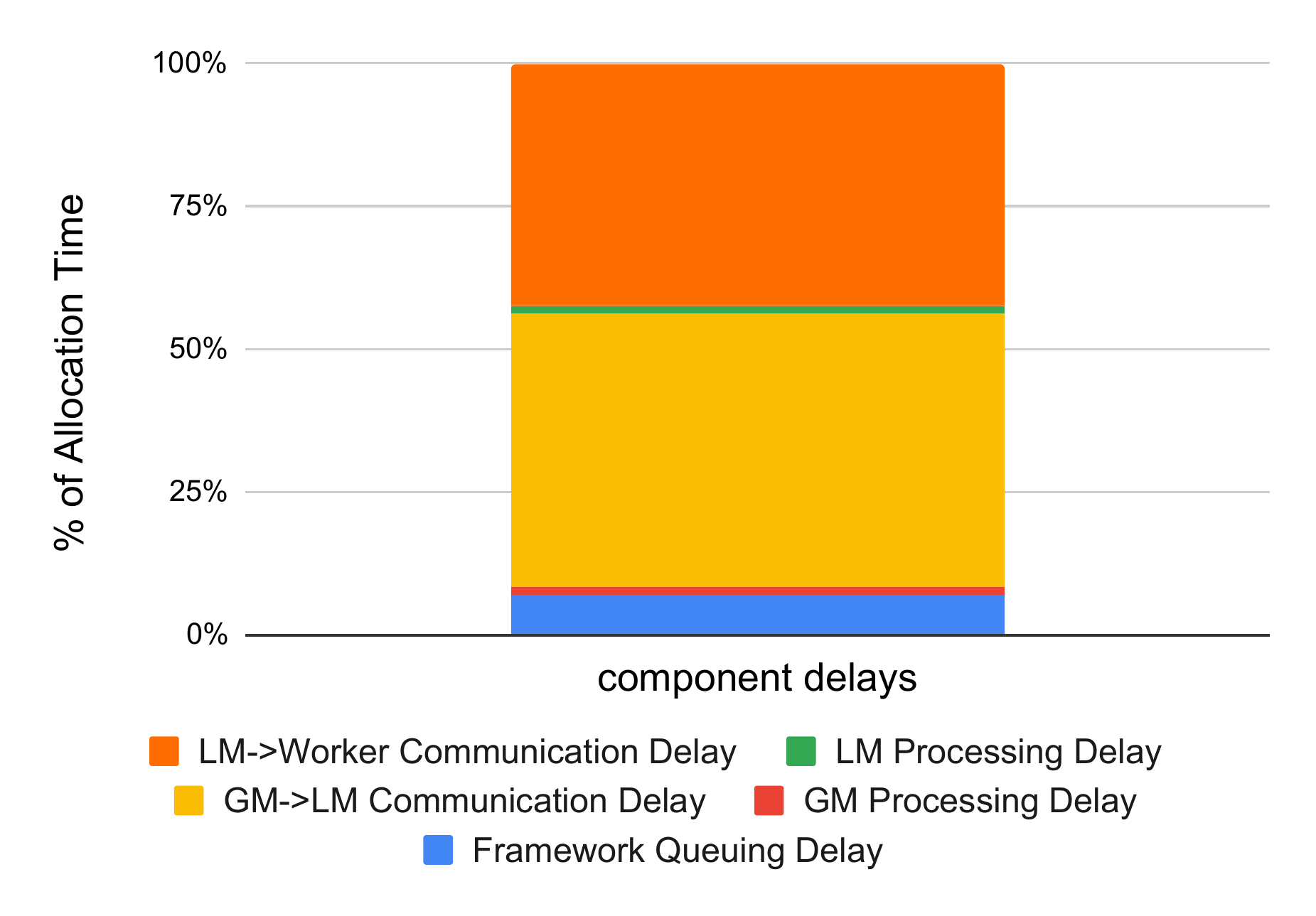}
    \caption{ Contributions of the Component Delays}
    \label{fig:components}
\end{figure}

We plotted Megha's component delays (Fig. \ref{fig:components}) described in Section \ref{sec:alloc_time} taking the median of the delays recorded for the simulation runs with the Cloudera trace. The results show that the largest contributor is the \(Communication\_Delay\) component which contributes to over 90\% of the delay. The \(Processing\_Delay\) components at the LMs and GMs contribute to less than 3\% of the allocation time. The \(Framework\_Queuing\_Delay\) constitutes $\sim$7\% of the allocation time.

\subsection{Cluster Size}
\begin{figure}[!b]
    \centering
       \includegraphics[width=\linewidth]{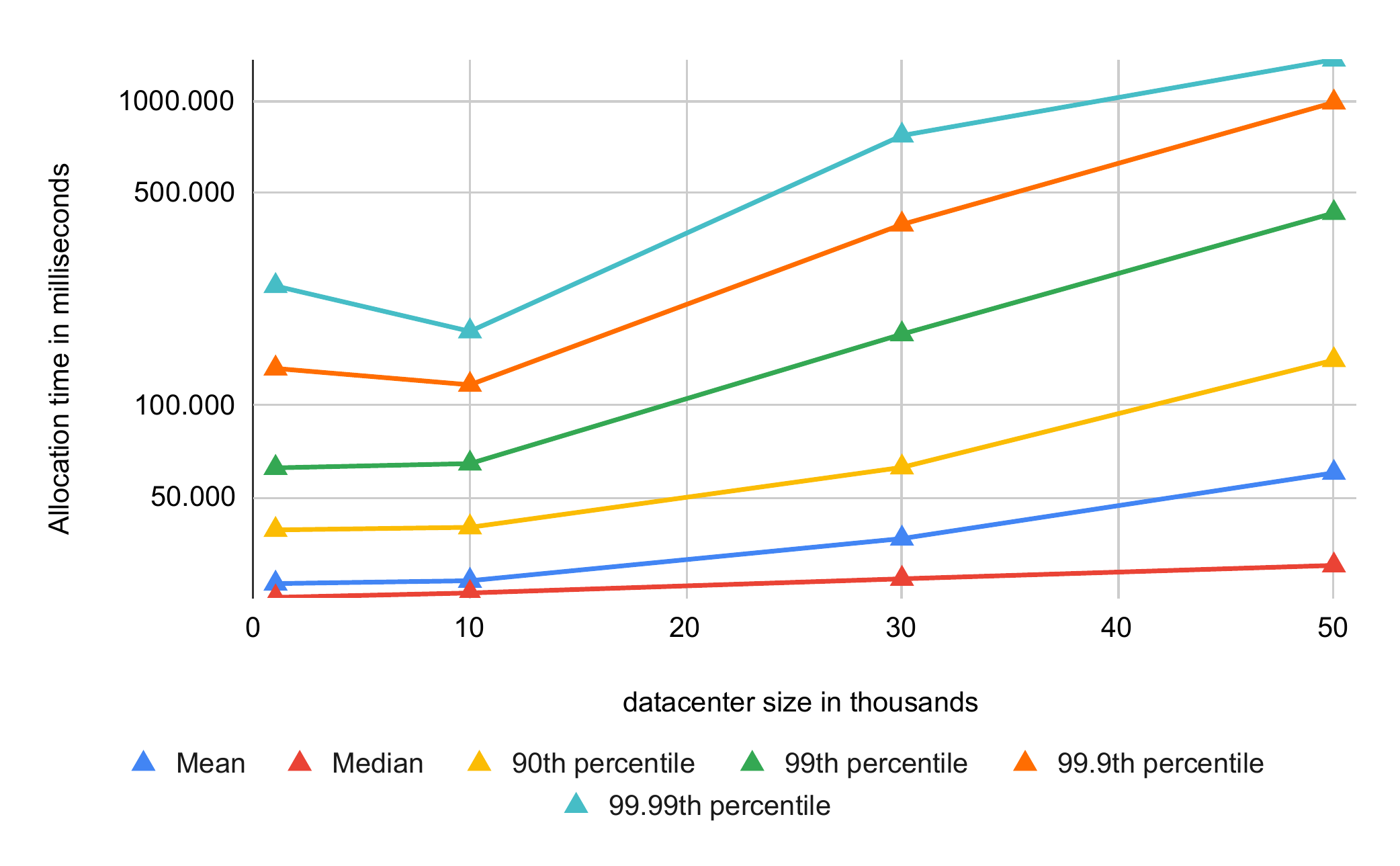}
        
    \caption{Allocation time recorded for different data center sizes with 10LM-10GM configuration}
    \label{fig:cluster_size}
\end{figure}

We study the impact of data center size on Megha's allocation time keeping the number of LMs and GMs constant at 10. Fig. \ref{fig:cluster_size}, shows how Megha's resource allocation time increases with an increase in the number of workers when the number of LMs and GMs are not correspondingly increased. This delay can be attributed to the increase in the amount of global state, and the size of each LM update that a GM must process as the number of workers increases.

\subsection{Number of LMs and GMs}

\begin{figure}[]
    \centering
       \includegraphics[width=\linewidth]{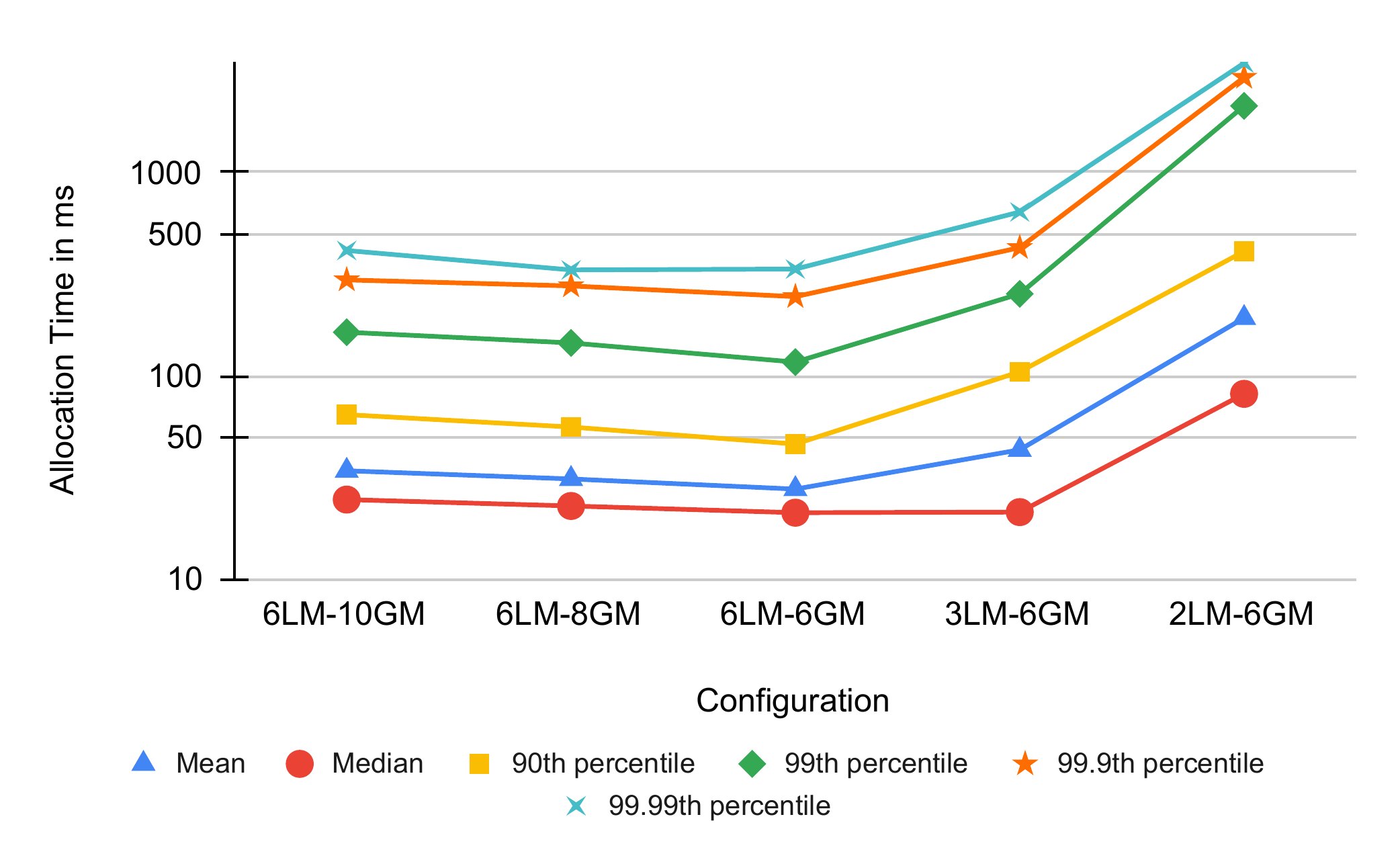}
        
    \caption{Allocation time recorded for different configurations}
    \label{fig:config}
\end{figure}

We study the impact of the number of LMs and GMs on the task allocation time, using simulations run with the \texttt{alibaba} trace for a data center of size 30,000 worker nodes (Fig. \ref{fig:config}. Increasing the number of GMs increases the scheduling throughput of the system, thereby decreasing the \(Framework\_Queueing\_Delay\) experienced by each task. Increasing the number of LMs increases the total number of GM requests that can be processed per second. It also reduces the size of the updates sent to the GMs, resulting in smaller processing overheads per request. However, when the number of GMs exceeds the number of LMs, we find that there is a slight increase in the allocation times reported. This can be attributed to the LMs being unable to keep up with the throughput of the GMs resulting in a larger \(Framework\_Queueing\_Delay\) at the LMs.

\subsection{Fairness}

\begin{figure}[]
    \centering
       \includegraphics[width=\linewidth]{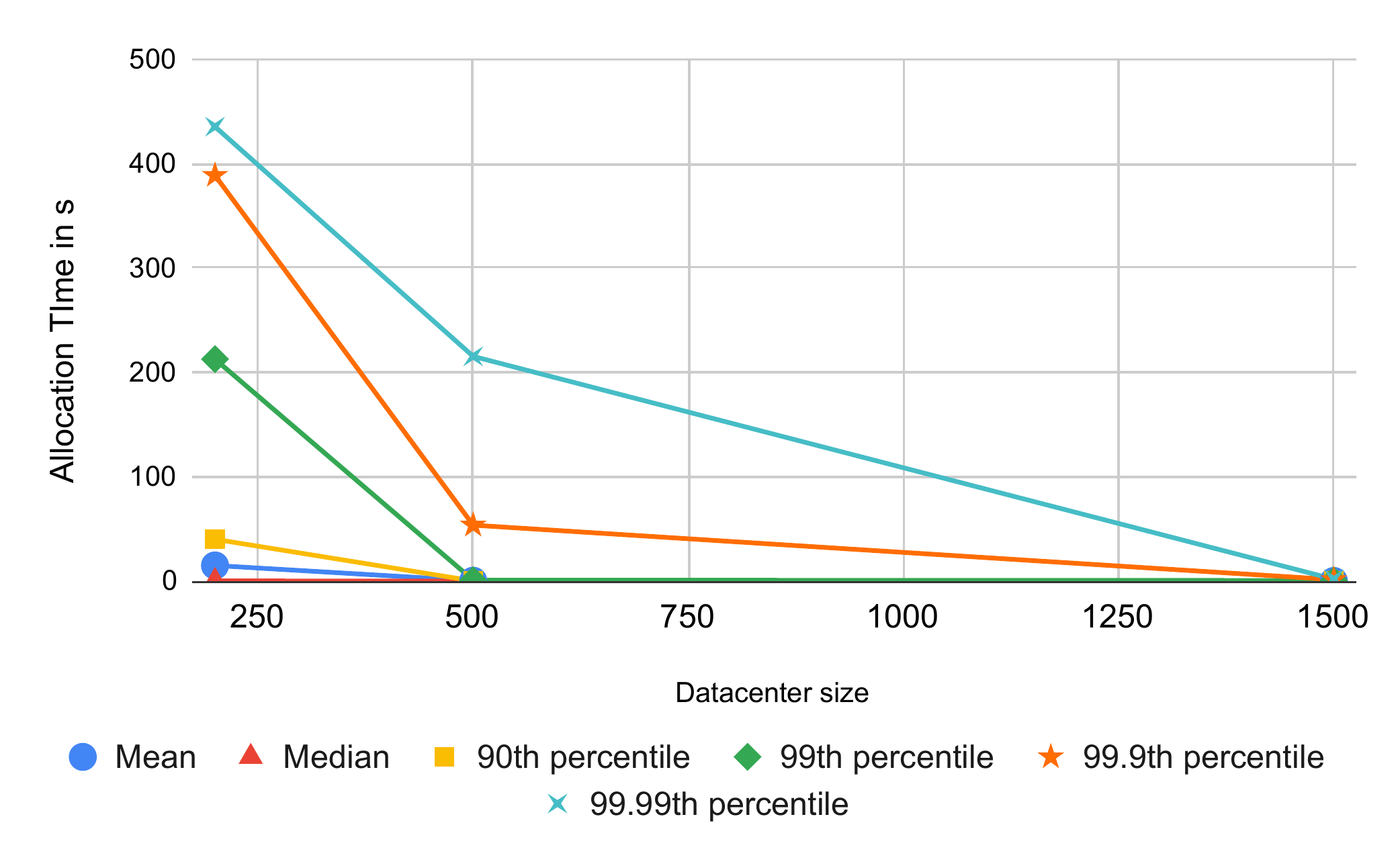}
        
    \caption{Allocation time recorded for different data center sizes with fairness}
    \label{fig:fair}
\end{figure}

\begin{figure}[]
    \centering
       \includegraphics[width=0.9\linewidth]{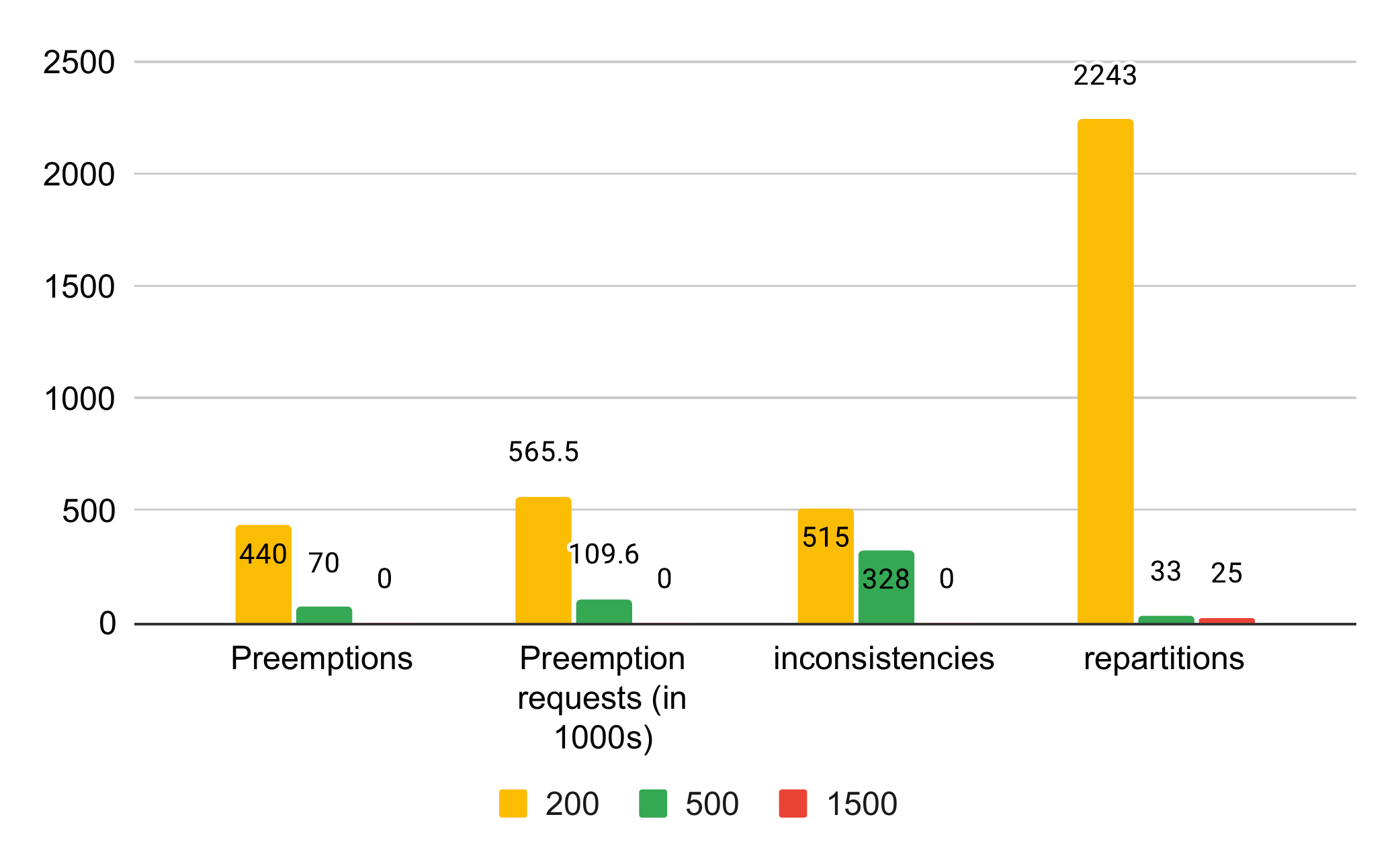}
        
    \caption{Number of preemptions, repartitions and preemption attempts recorded for each data center size}
    \label{fig:preempt}
\end{figure}

Megha's fairness algorithm only kicks in when there is resource contention amongst the users. Therefore, we chose smaller data center sizes to encourage resource contention. Fig. \ref{fig:fair} shows the allocation times of tasks when the simulation is run for data centers of sizes 200, 500, and 1500. We configured Megha with 4 user queues having 10\%, 25\%,15\%, and 50\% resource shares. The first user queue is mapped to GM 1, the second and third to GM 2, and the last to GM 3. We have used the \texttt{google\_19} trace for this analysis. The number of repartitions, preemption attempts, and task preemptions is shown in Fig. \ref{fig:preempt}. The allocation time is seen to increase with a decrease in the cluster size. This is because, with a decrease in the cluster size, resources are scarcer and tasks are made to wait longer for the required resources to free up. The number of preemptions, preemption attempts, and repartitions also follow the same trend and is found to increase with the decrease in cluster size.


\subsection{Performance Evaluation with Prototype}

\begin{figure}[!t]
    \centering

       \includegraphics[width=0.7\linewidth]{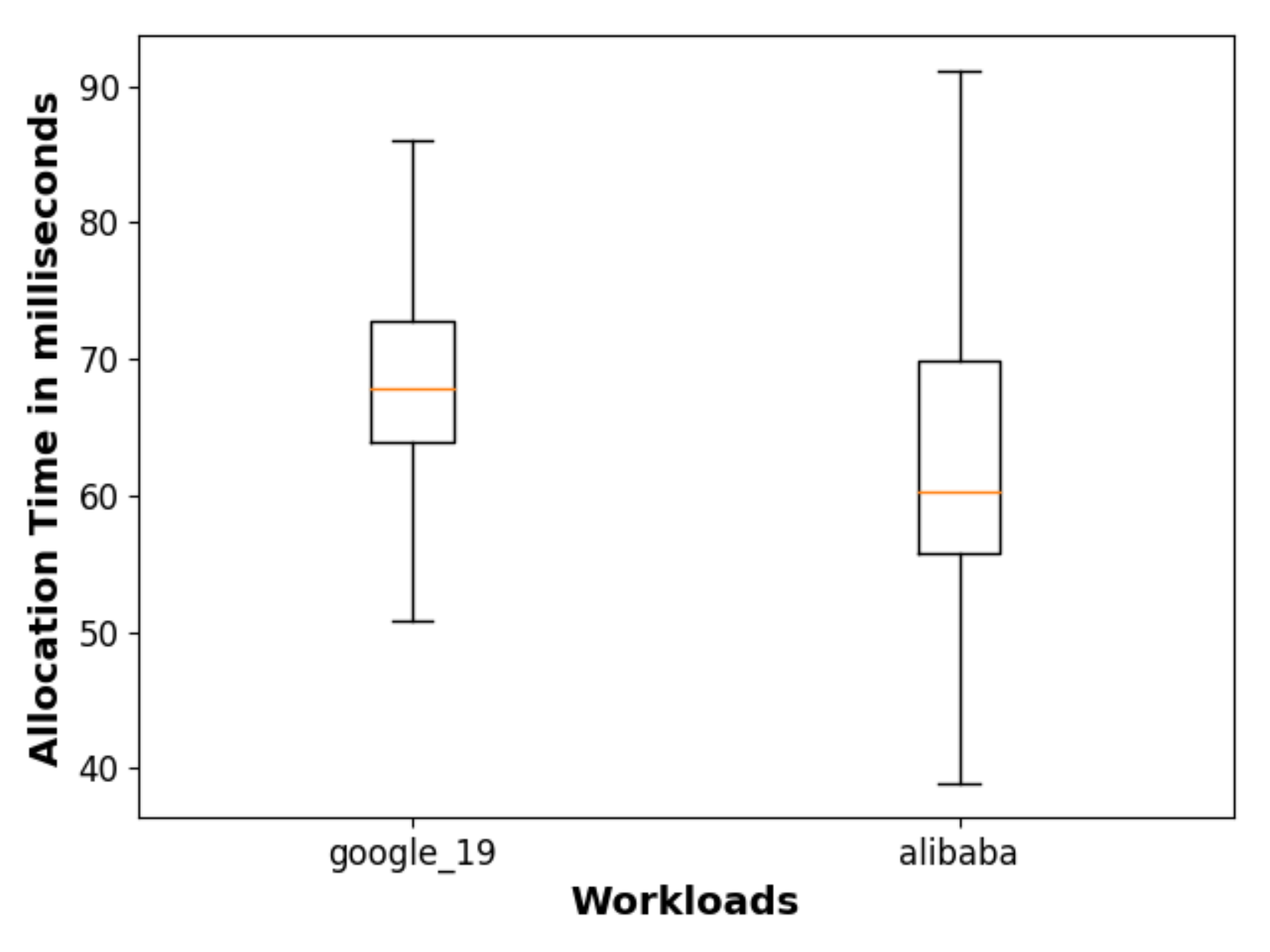}
   \caption{Distribution of the allocation time recorded for the workloads scheduled using the prototype}
    \label{fig:120_allocation}
\end{figure}

The distribution of the allocation time of the prototype evaluated on Linode instances is shown in Fig.~\ref{fig:120_allocation}. The allocation times recorded in the prototype were found to be higher than those recorded during the simulation. This can be explained by the delays observed in the tasks' execution on each worker node. The median execution delays recorded were 3.32 s for \texttt{google\_19} and 2.15s for \texttt{alibaba}. There is no execution delay recorded in the simulation results because the simulator does not account for interference. The prototype, on the other hand, runs the task such that the task completes its execution only after it finishes consuming the requested resources for the task's duration. That is, the task requests 0.5 CPU cores for 2 seconds, the task is said to have completed its execution only after consuming 1 CPU-core-second on the worker node. Due to interference on worker nodes with co-located tasks, the tasks need to run for longer than the duration specified to consume the required resources. This results in the resources being held for longer, which consequently leads to larger allocation times in tasks that require resources on the same worker node. The load from the google\_19 workload is higher, causing greater resource contention and interference thereby leading to larger delays in the task completion times.

\section{Conclusion}\label{sec:conclusion}
Data center resource management and scheduling frameworks have received a lot of attention in the recent past. The frameworks must be able to ensure optimal performance of workloads and at the same time, guarantee low-latency allocation decisions. This paper describes a decentralized resource management framework for federated clusters, named Megha. The framework uses a flexible partitioning scheme and eventual-consistency to achieve scalability and low allocation times while satisfying the various placement constraints of the tasks and guaranteeing global fairness. Our experiments conducted with workloads derived from 3 production traces, and with different data center sizes ranging from 120 to 50k nodes, show that Megha can achieve median allocation times comparable to that of distributed schedulers (Sparrow), and improves on their 99th percentile tail latency by 2 orders of magnitude.

\section{Future Work}\label{sec:future}
We are currently working on two extensions to the study presented in the paper. The first is to evaluate the impact of varying the heartbeat period on the overall performance of the system. We also wish to extend Megha to support scheduling policies such as gang scheduling, and priority-based scheduling.

% % use section* for acknowledgment
% \ifCLASSOPTIONcompsoc
%   % The Computer Society usually uses the plural form
%   \section*{Acknowledgments}
% \else
%   % regular IEEE prefers the singular form
%   \section*{Acknowledgment}
  
% \fi
% The authors would like to thank the following undergraduate students from the Department of Computer Science and Engineering, PES University, for their valuable assistance with the collection of trace samples, and the implementation of the slave modules --- Arindaam Mandal, Ishitha Agarwal, Kavya P K, Maddi Siddart, Neil John, Praveen Kumar, Rishith Bhowmick, Siva Raja Ganesh A, and Sonam Shenoy. 

\printbibliography
% that's all folks
\end{document}